\documentclass{elsart}

\usepackage{natbib}

 \usepackage{epsfig}

\usepackage{amssymb}
\usepackage{graphicx}

\begin{document}

\begin{frontmatter}

\title{High frequency intrinsic modes in El Ni$\tilde{n}$o/
Southern Oscillation Index}
\author{Filippo Petroni}
\ead{{\em Corresponding author } fpetroni@gmail.com}
\author{and Marcel Ausloos} 
\address{GRAPES, B5, Sart-Tilman, B-4000 Liege, Belgium}

\begin{abstract}
Recent $daily$ data of the Southern Oscillation Index have been analyzed. The power spectrum indicates major intrinsic geophysical short periods. 
  We find interesting ``high frequency'' oscillations at 24, 27, 37, 76, 100 and 365 days.
   In particular the 24 days peaks may correspond to the Branstator-Kushnir wave, the 27 days may be due to the moon effect rotation, the 37 days peaks is most probably related to the Madden and Julian Oscillation. It is not yet clear the explanations for the 76 days which may be associated with interseasonal oscillation in the tropical atmosphere; the  100 days could be resulting from a mere beat between the 37 and  27 periods,  or the 76 and 365 days.
Next these periods are used to reconstruct the signal and to produce a forecast for the next 9 months, at the time of writing. 
  After cleansing the signal of those periodicities a detrended fluctuation analysis is performed to reveal the nature of the stochastic structures in the signal and whether specific correlation can be found. We study the evolution of the distribution of first return times, in particular between $extreme$ $events$. A markedly significant difference from the expected distribution for uncorrelated events is found.
  \end{abstract}

\end{frontmatter}

\section{Introduction}
One of the most intriguing phenomena in climatology is  known as {\it El Ni$\tilde{n}$o/Southern Oscillation} (ENSO), i.e.
the more or less cyclic warming and cooling, mainly seen through an oscillation of about $6-8$ years,  of the eastern and central regions in
the Pacific Ocean. El Ni$\tilde{n}$o is considered to be due to a disruption of the ocean-atmospheric system in the tropical
Pacific and is
factually described by the so called {\it Southern Oscillation Index (SOI)} \cite{val88,phi99,ghi+98,kep+92}, a proxy measure  based on surface air pressure differences between Darwin, Australia and Tahiti, French Polynesia \cite{wal28}.
Data are usually analyzed from $monthly$ averages. 

Sustained negative values of the SOI often indicate El Ni$\tilde{n}$o episodes (see {\it http://www.bom.gov.au/climate/glossary/soi.shtml}).
These negative values are usually accompanied by sustained warming of the central
and eastern tropical Pacific Ocean, a decrease in the strength of the Pacific
Trade Winds, and a reduction in rainfall over eastern and northern Australia. The
most recent strong El Ni$\tilde{n}$o was in 1997/98. Positive values of the SOI
are associated with stronger Pacific trade winds and warmer sea temperatures to
the north of Australia, popularly known as a La Ni$\tilde{n}$a episode. Waters in
the central and eastern tropical Pacific Ocean become cooler during this time.
Together these give an increased probability that eastern and northern Australia
will be wetter than normal. The most recent strong La Ni$\tilde{n}$a was in
1988/89; a moderate La Ni$\tilde{n}$a event occurred in 1998/99, which weakened
back to neutral conditions before reforming for a shorter period in 1999/2000.
This last event finished in Autumn 2000.

In our opinion wide window filtering and averaging techniques are nowadays unnecessary to obtain interesting informations.   In fact,  one of the most important problems in quantitative weather forecasting is to understand the nature (or structure) of stochastic processes which underline the weather evolution.  One aspect of interest is to determine characteristics of the fluctuation distribution   in a signal  \cite{aus02}, leading to a probability distribution function (PDF) and their correlations. Empirical studies found that they are not (always) Gaussian distributed \cite{aus+01}.
It was recently found e.g. for the southern
oscillation index (SOI) \cite{aus+01} that long-range correlations may exist between the fluctuations of the index (even if it is still matter of debate \cite{met03}; moreover the PDF's have so called heavy tails (\cite{aus+07}), both features being describable through a Fokker-Planck equation approach. For such highly non linear systems higher frequency inputs of initial conditions  in numerical simulations should be clearly helpful. Computer time would also be reduced if scaling laws are found or known, and propose criteria/constraints in iterating models.

Here below we report results of  analysis of $daily$ data downloaded from the Long Paddock - Climate Management Information for Rural Australia web site {\it http://www.longpaddock.qld.gov.au/ SeasonalClimateOutlook/ SouthernOscillationIndex/ SOIDataFiles/ index.html} for the time interval between Jan. 1999 and March
2007. We look for intrinsic mode periodicity and short/long  term correlations present in the daily time series.  We find interesting ``high frequency'' oscillations at 24, 27, 37, 76, 100 and 365 days. For correlation studies we use the detrended fluctuation analysis (DFA) method, equivalent to finding  the Hurst exponent \cite{hur51,tur97}, on the {\it cleansed signal} (to be defined later).

Whence  we reconstruct the signal, extrapolate for forecasting-like purpose for the next 9 months, at the time of writing. The presently objective  observations seem to give some good agreement. We measure the so called area under the receiver operating curve as a test of forecasting quality \cite{sta+89,ste00}. We get a 0.73 value for the area, a value which is usually considered quite  good.

  Moreover considering the signal as a random walk we can estimate the law of ``first returns'' for barrier level crossing, giving some quantitative information on the distribution time intervals even between $extreme$ $events$. A markedly significant difference from the expected distribution for uncorrelated events is found. The dynamics of ENSO being sufficiently well represented in the daily data, we concur that a $daily$ time series may be quite usefully employed for predicting events.

\section{Data} 
\begin{figure}
\includegraphics[width=20pc]{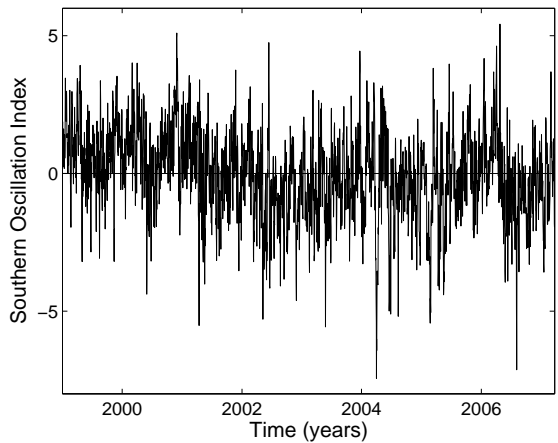} 
\caption{Daily values of the Southern Oscillation Index
(SOI) as defined in the text reported from 01 Jan. 1999 to 25 March 2007. Data are
downloaded from
$http://www.longpaddock.qld.gov.au/$ $SeasonalClimateOutlook/$ $SouthernOscillationIndex/$ $SOIDataFiles/index.html$.
Data series consists of 3006 data points.}\label{fig1}\end{figure}

Daily values of the $SOI$ index for the time interval between Jan. 1999 and March.
2007 were downloaded from the Long Paddock - Climate Management Information for Rural Australia web site {\it http://www.longpaddock.qld.gov.au/ SeasonalClimateOutlook/ SouthernOscillationIndex/ SOIDataFiles/ index.html} for the longest period available at the time of finishing this study,
(the website is updated daily). The raw data series is normalized through
the standard deviation $S$ of the sea level pressure (SLP) at a station in Tahiti
and the sea level pressure at a station in Darwin. 
\begin{equation} SOI = \frac{P_{Tahiti} - P_{Darwin}}{S} \end{equation}
The data set consists of $3006$ data points. The $SOI$ daily data are plotted as a function of time in Fig. \ref{fig1}. The mean and standard deviation of this time series are respectively $\mu = -0.05$ and $\sigma =1.57$.

It is sometimes stated that daily or weekly values of the SOI do not convey much
in the way of useful information about the current state of the climate, and
accordingly the Bureau of Meteorology does not issue them. Daily values in
particular can fluctuate markedly because of daily weather patterns, and should
not be used for climate purposes. We may disagree with this statement. There are
indeed techniques which can sort out noise from coherent behavior and conversely. This shows the intent of using daily data.

\section{Power spectrum}\label{PS}
\begin{figure}
{\bf a)} \vskip .01cm
\noindent\includegraphics[width=20pc]{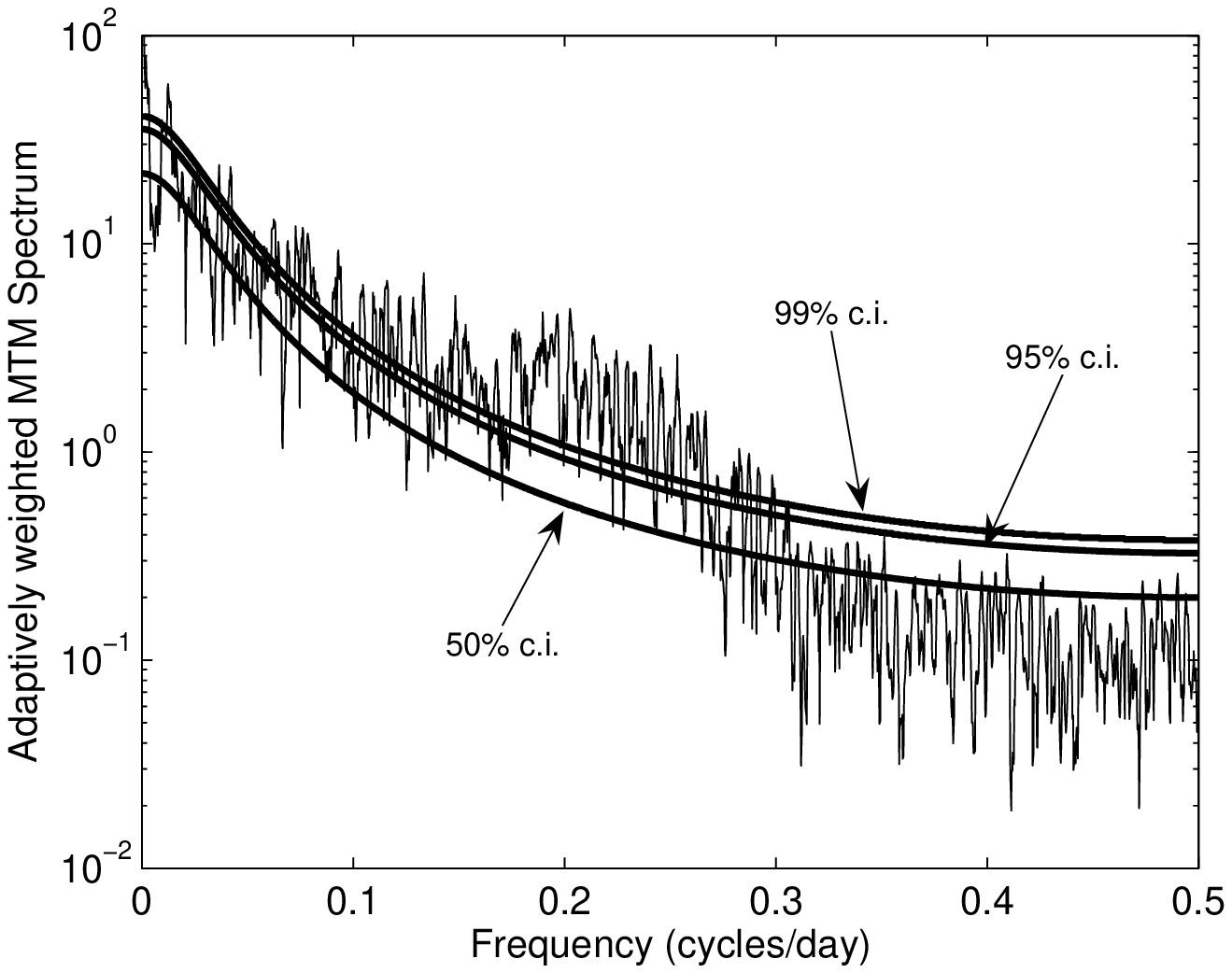}
\vskip .01cm {\bf b)} \vskip .01cm
\noindent\includegraphics[width=20pc]{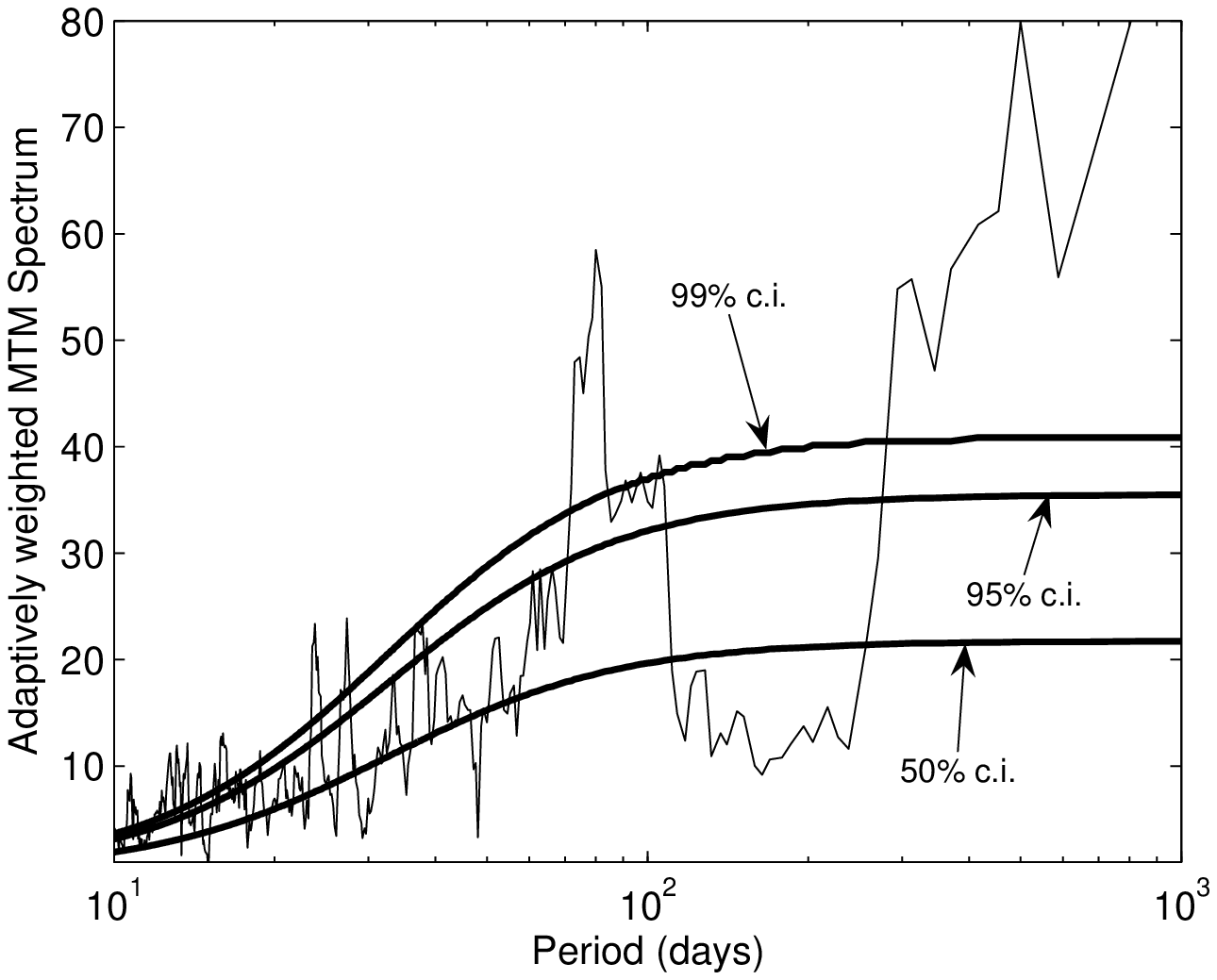}
\caption{a) Adaptively weighted multitaper power spectrum of SOI daily data as function of frequency. Continuous lines are red noise confidence interval. b) same as before but plotted as function of time (days) to highlight the periodicities in days of the SOI intrinsic high frequency modes. Continuous lines are red noise confidence interval. } \label{fig2}
\end{figure}

\begin{figure}
{\bf a)} \vskip .01cm
\noindent\includegraphics[width=20pc]{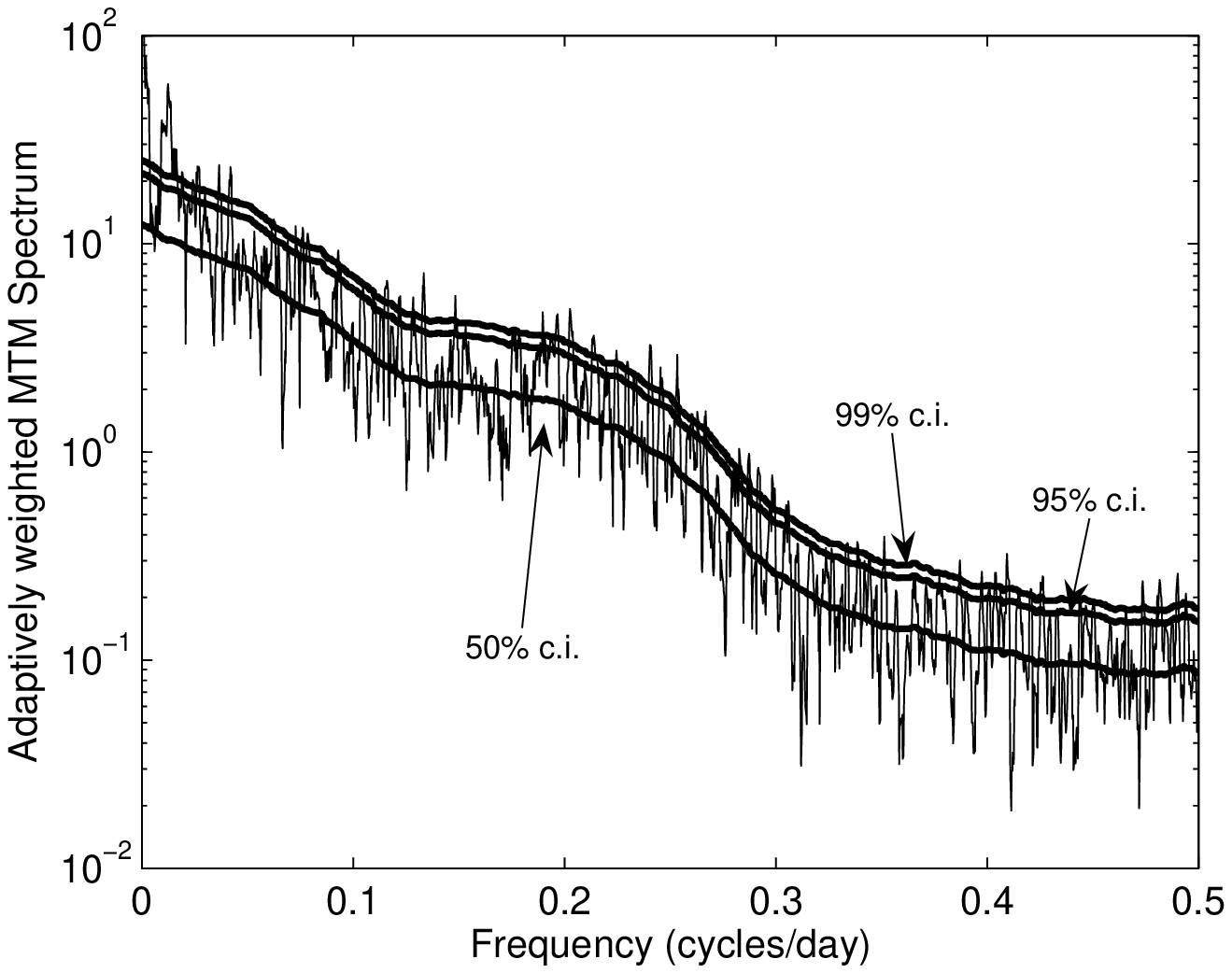}
\vskip .01cm {\bf b)} \vskip .01cm
\noindent\includegraphics[width=20pc]{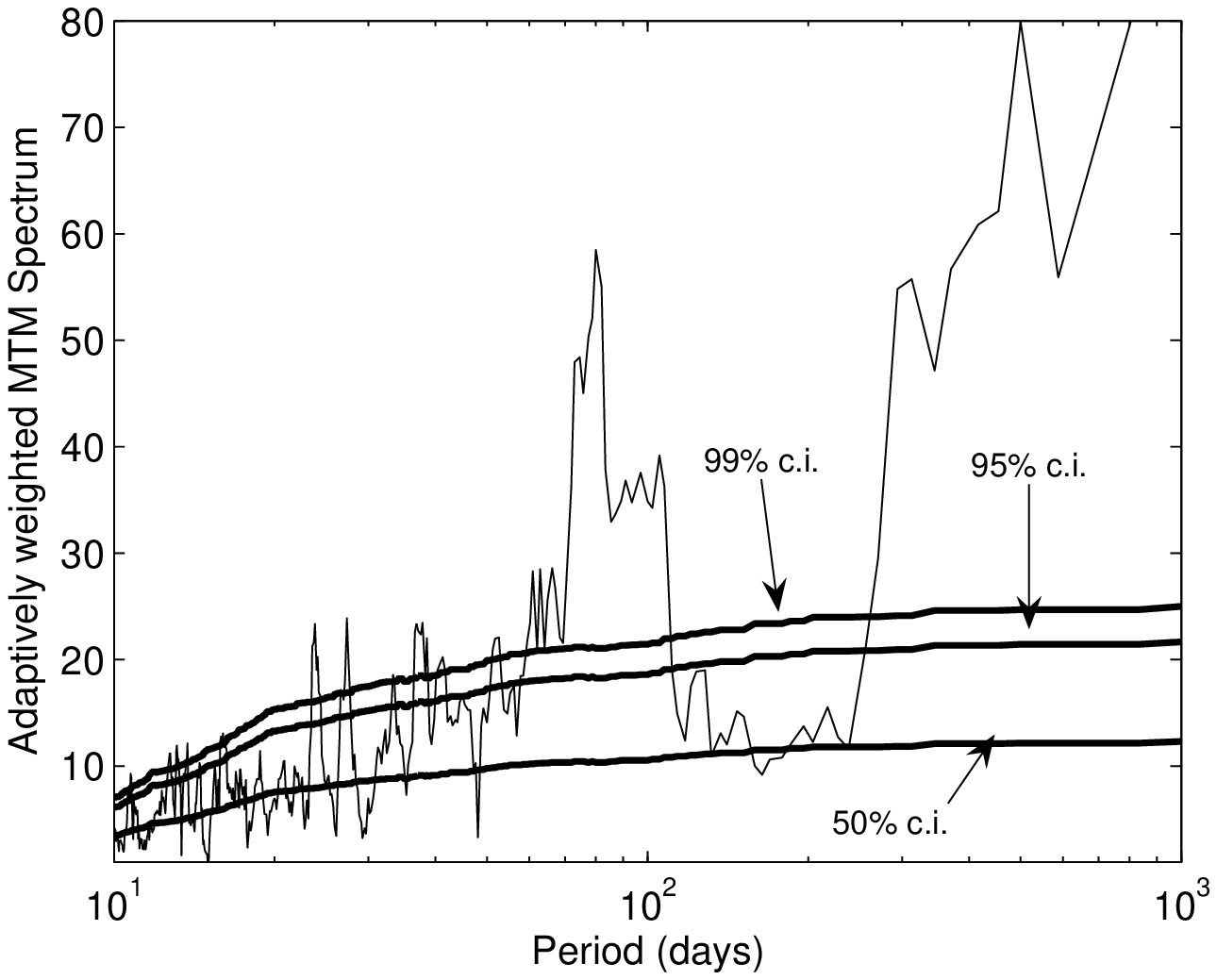}
\caption{a) Adaptively weighted multitaper power spectrum of SOI daily data as function of frequency. Continuous lines are locally white noise confidence interval. b) Same as before but plotted as function of time (days) to highlight the periodicities in days of the SOI intrinsic high frequency modes. Continuous lines are locally white noise confidence interval.} \label{fig3}
\end{figure}

Instead of empirically searching for modes \cite{sal+02} we directly estimate the intrinsic fast (intra annual) modes from the power spectrum (PS) of the daily data following the algorithm proposed by \cite{man+96}. The PS is estimated using the Multitaper Method (MTM) \cite{tho82,per+93} with $K=5$ tapers. We chose this method because it is non parametric and reduces the variance of spectral estimates by using a small set of tapers \cite{tho82,per+93}.
The PS obtained through MTM is shown in Fig. \ref{fig2}a on a linear-log plot. The PS of daily SOI is tested for significance relative to the null hypothesis of red noise (Fig. \ref{fig2}a) and locally white noise (Fig. \ref{fig3}a). A robust estimate of the red noise spectral background is found, following the work by \cite{man+96}, by minimizing the misfit between an analytical AR(1) (Auto Regressive process of order 1) red noise spectrum given by
\begin{equation}\label{ar1} S(f)= S_0 \frac{1-r^2}{1-2r\cos(2\pi f/f_N)+r^2} \end{equation}
and the adaptively weighted multitaper spectrum which has been firstly convolved with a median smoother to reduce the weight of outliers in the least square fit. In  equation (\ref{ar1}) $S_0$ is the average value of the power spectrum, $r$ is the lag-one autocorrelation and $f_N=2/\Delta t$ is the Nyquist frequency ($\Delta t= 1$ day), the highest frequency that can be resolved with a sampling rate $\Delta t$. In the robust estimation $S_0$ and $r$ are considered as free parameters for the minimization procedure. The PS is also tested against a locally white noise hypothesis (Fig. \ref{fig3}a).

The 95\% and 99\% confidence level (showed in the plots) for peaks detection are determined from the appropriate quantile of the $\chi$-square distribution with $\nu =2K$ degrees of freedom. From the plots we can say that the red noise background hypothesis is appropriate only for the first part of the spectrum (small frequency). In the frequency range $[0;0.1] Hz$ both hypotheses on background noise give similar results on the number and positions of peaks exceeding the 99\% confidence intervals. Quite different are instead the results for higher frequencies for which the spectrum of the SOI signal does not seem to be well fitted by a red noise spectrum.    

From figures \ref{fig2}b and \ref{fig3}b one can recognize the position of several peaks at specific days (or for well defined periods) in the power spectrum. The positions of the highest peaks exceeding the 99\% confidence interval are found to be roughly at 24, 27, 37, 76, 100 and 365 days; we did not consider those peaks which had small power even if exceeding the 99\% confidence level, i.e. for frequencies greater than 0.1 cycles/days. Moreover we realize that each is an integer value, which is a first rough approximation with respect to the true geophysical period representing some event. Yet, some of these periodicities may be related to well known geophysical processes \cite{kon+04}. In particular the 24 days peaks may correspond to the Branstator-Kushnir wave, a westward travelling wave with a period of about 23 days \cite{brs87,brs92,brs+95,sim+83,kus87}, the 27 days may be due to the moon effect rotation, the 37 days peaks is most probably related to the extratropical wave, so called Madden and Julian Oscillation (MJO) \cite{mar+96,ghi+91}. It is not yet clear which explanations can be provided for the 76 days which may be associated with interseasonal oscillation in the tropical atmosphere, as discussed by \cite{ghi+02} analyzing monthly SOI data; the  100 days could be resulting from a mere beat between the 37 and  27 periods,  or the 76 and 365 days, which give respectively a 100 and a 96 day beating period, - or both jointly;  the 365 days periodicity nature is obvious for those who regularly  turn around the sun.

\subsection{Signal reconstruction and forecasting}
\begin{figure} 
\noindent\includegraphics[width=20pc]{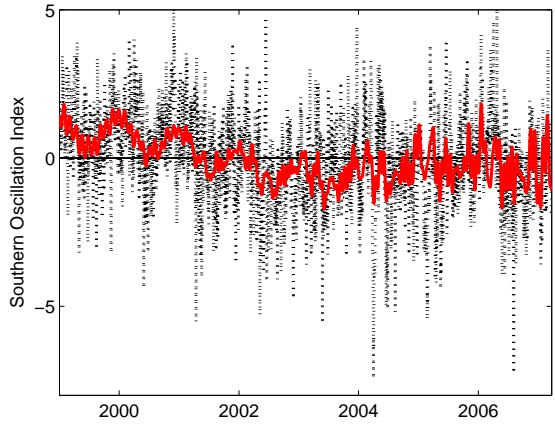} 
\caption{Daily SOI signal and its reconstruction with basic periods as indicated in the text.}\label{fig4} \end{figure}

The results of the previous section can be used to reconstruct and forecast the future behavior of the daily SOI signal. 
We assume that the SOI signal can be modeled in time by the following stochastic process
\begin{equation}\label{model} SOI(t) =A_1(t) + \sum A_i(t)\cos (\frac{2\pi}{T_i}t +\phi_i) + \epsilon (t)\end{equation}
where $\epsilon (t)$ is a noise component, the nature of which should be determined, and $T_i$ are the periods obtained in the previous analysis. The values for $A_i(t)$ and $\phi_i$ can be obtained from a time domain inversion of the spectral domain information contained in the $K$ eigentapers \cite{par92,man+94} but we prefer to model $A_i(t)$ as polynomial functions of degree $n$ (in our case we consider $n=1$, i.e. $A_i(t)=A_i + B_it$) and to obtain the free parameters from a least square fitting procedure.
In Fig. \ref{fig4} we show the signal (dashed line) and the reconstructing function (solid line). The periods used for reconstruction are the following (the parameters $A_i$, $B_i$ and $\phi_i$ obtained by the best fit procedure are summarized in Table 1): 70 months \cite{aus+07}, 365 days, 100 days, 76 days, 37 days, 27 days and 24 days. 

\begin{figure}
\noindent\includegraphics[width=20pc]{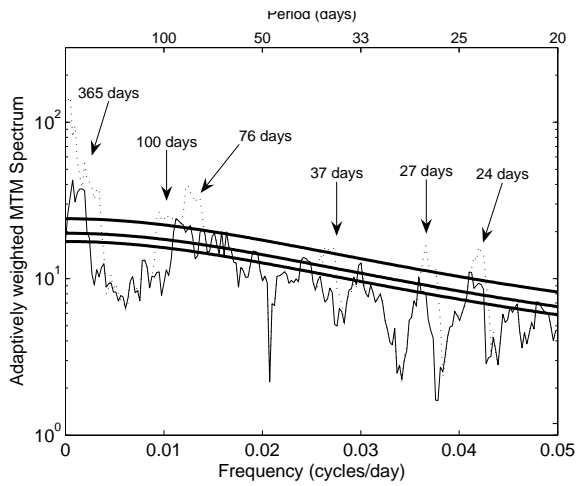} 
\caption{Discrepancy between the adaptively weighted multitaper power spectrum of daily SOI (dashed line) and cleansed signal (solid line). The solid thicker lines represent the 90\%, 95\% and 99\% confidence interval as explained in the text.} \label{fig5}
\end{figure}

The function so obtained is then removed from the original SOI time series.  A cleansed signal is obtained and the MTM spectrum is estimated again. In Fig. \ref{fig5} we show a comparison of the power spectrum before and after removing the harmonic components from the daily SOI. One can recognize the portion of the spectrum due to the harmonic components. Both spectra are compared with the 95\% and 99\% confidence levels as described above for the red noise hypothesis. Recall that one relevant erroneous ingredient is the assumption of integer values for the considered geophysical periods. The error may accumulate for moderately short time series, but this assumption is hardly unavoidable.

\begin{figure}
\noindent\includegraphics[width=20pc]{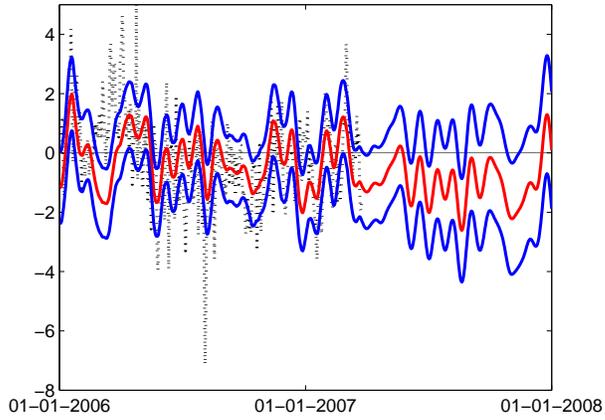} 
\caption{Daily SOI (dashed line) together with the prediction (solid lines) based on the function described in the text. The upper solid and lower solid lines represent the 95\% confidence interval estimated throught a cross validation procedure (see text for details).} \label{fig6}
\end{figure}

In Fig. \ref{fig6} we show the same reconstruction function from 01 Jan. 2006 and a possible forecast, based on this function, for the SOI for the next year (after March 2007 until 01 Jan. 2008). This function is shown together with a 95\% confidence interval obtained through a cross validation procedure where each new data point is predicted using a model fitted only on the previous data points (so called ``leave one out'' validation).

\begin{figure} 
\noindent\includegraphics[width=20pc]{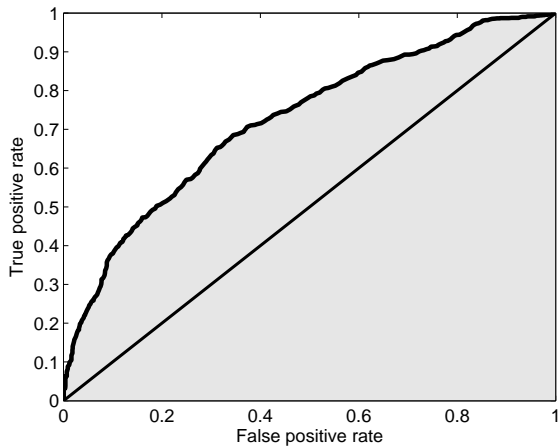} 
\caption{Receiver operating curve (ROC), -thick line compared to random estimates, -straight line. The shadowed area represent the ROC AUC (see text) } \label{fig7}
\end{figure}

In order to give another evaluation of the performance of our forecasting algorithm in predicting positive and negative SOI values we have also estimated the receiver operating curve (ROC) (plot in Fig. \ref{fig7}). The ROC \cite{sta+89,ste00} is a widely used method to asses the validity of a binary classifier through the sensitivity and specificity as its discrimination threshold is varied. 
It is obtained as follows: in the first step SOI is digitalized based on the sign, the  forecasted time series is similarly translated according to the signs (- and +). We then verify how good the forecast is by estimating the true positive and false positive rates (as shown in Figure \ref{fig7}), in the whole time window, varying the threshold used to discriminate between positive and negative values. The last step consists in estimating the area under the ROC, often called ROC AUC for which we obtained a value of $0.73$.

\section{Time correlations} 
Thereafter we can look for correlations in the cleansed signals, in which the main trends have been removed, through some estimator. This is important to understand the nature of the stochastic component in equation (\ref{model}). Different estimators can be found in the literature for the long and/or short range dependence of fluctuations correlations \cite{taq+95,bro+91}, one of the most precise and/or common is the detrended fluctuation analysis (DFA) method, see e.g. \cite{peng+94,bun+00,hu+01}. The method has
been used previously to identify whether long range correlations exist in
non-stationary signals, in many research fields such as e.g. finance \cite{van+97,van+98},
cardiac dynamics \cite{iva+96} and of course meteorology \cite{iva+00,iva+99,bun+98,perl+99}, and by so many others that no exhaustive  list can be here given.  For
an extensive list of references see \cite{hu+01}. Briefly, we recall that the signal time series
$y(t)$ is $first$ $integrated$, to `mimic' a random walk $Y(t)$. The time axis (from 1 to $N$)
is next divided into non-overlapping boxes of equal size $n$; one
looks thereafter for the best (polynomial, of degree $m$) trend, $z_{n,m}$, in each box, and calculates the root mean square deviation of the (integrated) signal with respect to $z_{n,m}$ in each box. The average of such values is taken at fixed box size $n$ in order
to obtain
\begin{equation} F_m(n) = \sqrt{{1 \over N } {\sum_{i=1}^{N}
{\left[Y(i)-z_{n,m}(i)\right]}^2}} \end{equation} 
\noindent The box size is next varied over all possible $n$ values and $F_m(n)$ recalculated. The resulting function is expected to behave like $n^{\alpha}$ indicating a scaling law. The value of $\alpha$ can be related to the fractal dimension and/or the Hurst exponent of the signal \cite{tur97,aus02}.

\begin{figure} 
\noindent\includegraphics[width=20pc]{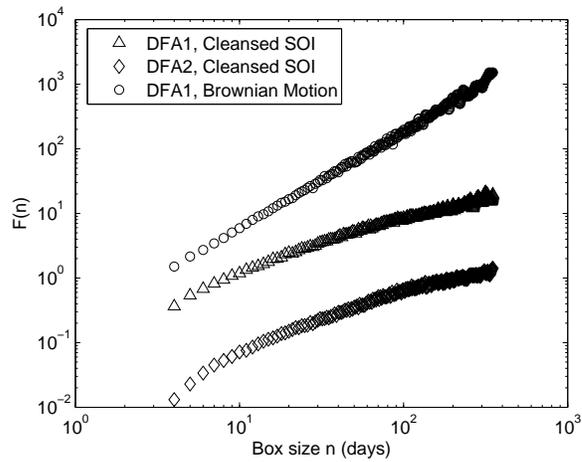} 
\caption{DFA1 (triangles) and DFA2 (diamonds, amplitude divided by 10) of cleansed SOI compared with DFA1 (circles, amplitude multiplied by 10) of a Brownian motion with the same data points (3006). The slope $\alpha$ for Brownian motion is $1.5$.} \label{fig8}
\end{figure}

\begin{figure} 
\noindent\includegraphics[width=20pc]{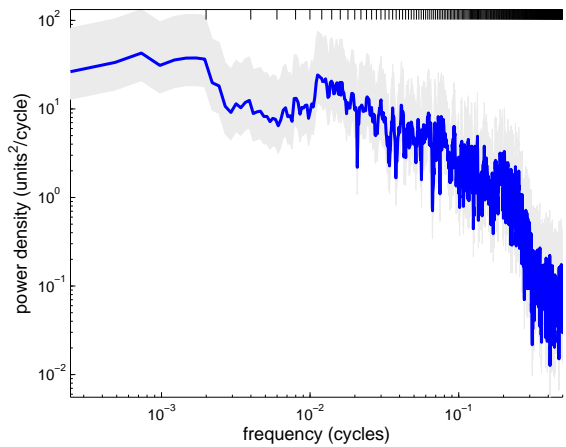} 
\caption{Log-log plot of the adaptively weighted multitaper power spectrum of cleansed SOI daily data as function of frequency. Gray shadow indicates 95\% confidence intervals.} \label{fig9}
\end{figure}

In Fig. \ref{fig8} we show a detrended fluctuation analysis of the cleansed daily SOI for linear ($m=1$, DFA1) and quadratic ($m=2$, DFA2) detrending. The resulting $F_m(n)$ functions can be usefully compared, from our point of view, with DFA1 of a synthetically generated Brownian motion with the same number of data points. It can be noticed looking at Fig \ref{fig8} that $F_m(n)$ for the daily SOI, for both detrending, does not show a linear behavior for the whole range of $n$ in the log-log plot, while the Brownian motion does. A not linear $F_m(n)$ function for the whole range of $n$ may be due to the process not being long memory or to trends \cite{hu+01} still present in the cleansed data and which have not been removed from the algorithm used here, an open question therefore remains. Another method to asses long memory in stochastic processes is through the PS \cite{mal+99}. We then show the adaptively weighted multitaper spectrum ($S(f)$) of the cleansed series (Fig. \ref{fig9}) in a log-log plot. Also in this case a linear behavior, which would imply $S(f)\approx f^{-\beta}$, is not found in the whole frequency range indicating no simple self-affine behavior.

\section{Return times} 
\begin{figure} 
\noindent\includegraphics[width=20pc]{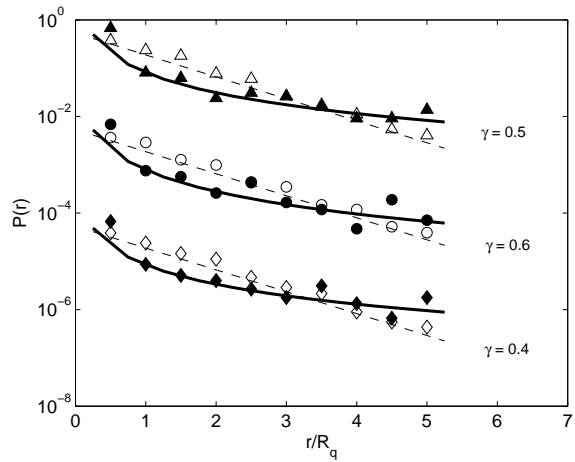} 
\caption{Return interval distribution for events exceeding negative (diamonds), positive (circles) and both positive and negative (triangles) threshold with $q= 1$ for cleansed SOI data (filled symbols) and shuffled cleansed SOI data (open symbols). The two type of distributions are fitted respectively with a stretched exponential (solid line) and a Poisson distribution (dashed line). The values for $\gamma$ are obtained through a least square fit.} \label{fig10}
\end{figure}

Another interesting point is the statistics of extreme (rare) events, namely those events that exceed a given threshold. Following previous studies we considered the cleansed SOI signal as the analog of a complex random walk and examine the distribution of first return times \cite{cim+99}. 
We studied the time interval $r$ between two consecutive threshold crossing.  The threshold $q$ measures the strength of an event. The time series is firstly normalized such to have variance equal to 1, then a threshold $q=1$ is chosen. We chose the same threshold for positive and negative ``extreme'' values. The results in Fig. \ref{fig10} show the distributions for the time interval between consecutive positive, negative and both positive and negative threshold crossing. 
The return time interval distributions for cleansed SOI are compared with those of shuffled SOI. It is expected \cite{bun+05,eic+06} that while the latter should follow a Poisson distribution, as for uncorrelated events, long range correlated process should follow a stretched exponential distribution \cite{bun+05,eic+06}. The two kind of experimental distributions have been fitted with a Poisson distribution, for the shuffled SOI, and with a stretched exponential distribution, defined as
\begin{equation}\label{exp}
P(r) = \frac{a}{R} e^{-b(r/R)^\gamma}
\end{equation}
for the cleansed SOI. In equation (\ref{exp}) $R$ is the mean value of $r$, $a$, $b$ and $\gamma$ are free parameters. The exponent $\gamma$, for a long memory process $x$, is the same exponent that characterizes the autocorrelation function at lag $s, \ C_x(s) = <x_ix_{i+s}> \approx s^{-\gamma}$.
Figure \ref{fig10} shows that while the shuffled data are well fitted by a Poisson distribution the cleansed SOI does not follow a Poisson distribution and it is closer to a stretched exponential one even if, due to the shortage of data, this result cannot yet be asserted with a good statistical significance test.

\section{Conclusion}
In contrast with many other works on SOI we have considered daily fluctuations instead of the more often used monthly averaged data. We have searched therefore for high frequency features,  and observed  intra annual major periods :  24, 27, 37, 76, 100 and 365 days. At periods shorter than 100 days, there is known evidence for very energetic westward propagating sea level signals at low latitudes (equatorward of about 20 deg latitude). The nature of this intraseasonal variability is usually thought to be distinctly different for periods longer and shorter than ca. 50 days. 
We have pointed out  that each period is an integer value, which is a first rough approximation with respect to the true geophysical period representing some event. Yet, some of these periodicities may be related to well known geophysical processes. The 24 days  may correspond to the Branstator-Kushnir wave, a westward travelling wave with a period of about 23 days \cite{brs87,brs92,brs+95,sim+83,kus87}, the 27 days may be tides due to the moon, the 37 days peaks is most probably related to the Madden - Julian Oscillation \cite{mar+96,ghi+91}. It is not yet clear which explanations can be provided for the 76 days which may be associated with interseasonal oscillation in the tropical atmosphere \cite{ghi+02}; the  100 days could be resulting from the beat between the 37 and  27 periods,  or the 76 and 365 days, which give respectively a 100 and a 96 day beating period, - or both jointly.

These periods have been used, by modeling the SOI as a quasi-oscillatory signal plus ``noise'', to reconstruct the time series and extrapolate it for forecasting purpose.
The nature of the ``noise'' is also studied by removing from the SOI time series the reconstructed time series. Detrended fluctuation analysis together with power spectrum analysis have been employed in the attempt to understand the nature of the stochastic process showing that no simple scaling can be assessed. 
Through a systematic study of  the distribution of first return times, we have indicated that extreme events cannot be considered uncorrelated.

This paper  has not primarily aimed at writing up a model nor interpreting the intrinsic modes, on one hand, - this is found in already published work, but rather in pointing out that daily, $quasi$ $unaveraged$ data can be used for medium range and high frequency forecasting.

\section*{Acknowledgements}
Part of FP work  has been supported by European Commission Project  E2C2 FP6-2003-NEST-Path-012975  Extreme Events: Causes and Consequences.
 Critical and encouraging comments by  B. Malamud, M. Ghil, P. Yiou, A. Witt and G. Rotundo  have been  very valuable for improving this report. Part of this work is related to activities in the COST P10 ``Physics of Risk'' program.

\newpage
\begin{table} \caption{Values of the parameter used for forecasting with a superposition of quasi-oscillatory signals.}
\begin{center} \begin{tabular}{|cccc|} \hline $T$ & $A$ & $B$ &  $\phi$ \\ \hline  70 months &-0.45 & 0 & 2.3 \\ 365 days & 0.58 & -0.0001 & 0.12 \\  100 days & 0.004 & 0.0002 & 1.4\\  77 days & -0.028 &  -0.0002 & 1.1 \\ 37 days & 0.38 & -0.0003 & 1.5 \\ 27 days & -0.084 & 0.0002 & 2.3 \\ 24 days & -0.008 & 0.0001 & 2.5 \\linear term & 0.56 & -0.0005 &  \\
\hline \end{tabular} \end{center} \label{table2} \end{table}

\newpage

\end{document}